\documentclass{ws-procs975x65}
\usepackage[utf8]{inputenc}
\usepackage{amsmath}
\usepackage{amsfonts}
\usepackage{color}

\def\beq{\begin{equation}}
\def\eeq{\end{equation}}

\begin{document}

\title{Hamiltonian Analysis In New General Relativity}
\author{Daniel Blixt$^{1,*}$, Manuel Hohmann$^1$, Martin Kr\v{s}\v{s}ák$^{1,2}$ and Christian Pfeifer$^1$}

\address{$^1$Laboratory of Theoretical Physics, Institute of Physics, University of Tartu,\\
Tartu, 50411, Estonia\\
$^2$Center for Gravitation and Cosmology, College of Physical Science and Technology, Yangzhou University,\\
Yangzhou 225009, China \\
$^*$E-mail: blixt@ut.ee\\
http://kodu.ut.ee/\textasciitilde  blixt/}


\begin{abstract}
It is known that one can formulate an action in teleparallel gravity which is equivalent to general relativity, up to a boundary term. In this geometry we have vanishing curvature, and non-vanishing torsion. The action is constructed by three different contractions of torsion with specific coefficients. By allowing these coefficients to be arbitrary we get the theory which is called ``new general relativity''. In this note, the Lagrangian for new general relativity is written down in ADM-variables. In order to write down the Hamiltonian we need to invert the velocities to canonical variables. However, the inversion depends on the specific combination of constraints satisfied by the theory (which depends on the coefficients in the Lagrangian). It is found that one can combine these constraints in 9 different ways to obtain non-trivial theories, each with a different inversion formula.
\end{abstract}


\bodymatter


\section{Conventions}
Greek indices denote global coordinate indices running from 0 to 3, small Latin indices are spatial coordinate indices running from 1 to 3, whereas capital Latin indices denote Lorentz indices running from 0 to 3. 
We are always dealing with Lorentzian metrics. Sign convention for the Minkowski metric is $\eta_{AB}=\mathrm{diag}(-1,1,1,1)$.

%
%
%
%
%
%
%
%
%

\section{Introduction}

%
%
%
%
%
%
%
%
%

Gravity is conventionally described with the Levi-Civita connection which is induced by a pseudo-Riemannian metric. This means that the covariant derivative of the metric is zero, and the connection is torsion-free but has curvature. However, there are equivalent theories to general relativity\cite{BeltranJimenez:2019tjy}. We will focus on teleparallel gravity\cite{Krssak:2018ywd} where we have vanishing curvature, but non-vanishing torsion.
\\ \indent In particular we will perform the Hamiltonian analysis of ``new general relativity'' (NGR)\footnote{With NGR, we refer to the more general three-parameter teleparallel gravity in contrast to the special one-parameter teleparallel gravity theory which NGR originally referred to\cite{Hayashi:1979qx}.}. For discussions of certain issues with these theories see\cite{Kopczynski,Nester1988,Cheng:1988zg} Previous work on the Hamiltonian analysis on teleparallel gravity theories have been performed in  \cite{Blixt:2018znp,Blixt:2019mkt,Blagojevic:2000qs,Li:2011rn,Ferraro:2018tpu,Nester:2017wau,Maluf:2001rg,Maluf:1994ji,Okolow:2013lwa,Okolow:2011nq,Okolow:2011np,Ferraro:2016wht,Cheng:1988zg}. However, the full Hamiltonian analysis of NGR has not been performed. NGR is described by the following action:
\begin{align}
	S_{\mathrm{NGR}}= m^{2}_{Pl}\int |\theta| \left(a_{1}T^{\mu}_{\ \nu\rho}T_{\mu}^{\ \nu \rho}+a_{2}T^{\mu}_{\ \nu \rho}T_{ \  \ \mu}^{ \rho  \nu}+a_{3}T^{\mu}_{\ \rho \mu}T^{\nu \rho}_{\ \ \nu} \right)\mathrm{d}^{4}x,
\end{align}
where $m_{Pl}$ is the Planck mass, $T^{\mu}_{\ \nu \rho}=\Gamma^{\mu}{}_{\rho\nu}-\Gamma^{\mu}{}_{\nu\rho}$
is the torsion component with $\Gamma^{\mu}_{\ \nu \rho}=e^{\ \mu}_{A}\partial_{\rho}\theta^{A}_{\ \nu}+e^{\ \mu}_{A}\left(\Lambda^{-1} \right)^{A}_{\ D}\partial_{\rho}\Lambda^{D}_{\ B}\theta^{B}_{\ \nu}$,
with $\theta$ being the tetrad, $e$ its inverse and $\Lambda$ is a Lorentz matrix. Global spacetime indices are raised and lowered with $g_{\mu\nu}=\theta^{A}_{\ \mu}\theta^{B}_{\ \nu}\eta_{AB}$, while Lorentz indices are raised and lowered with $\eta_{AB}$. A theory equivalent to general relativity is obtained by setting $a_{1}=\tfrac{1}{4}, \ a_{2}=\tfrac{1}{2}, \  \mathrm{and} \  a_{3}=-1$.

Alternatively, the NGR action can be written down in the so-called axial, vector, and tensor decomposition  \cite{Bahamonde:2017wwk}. Then
\begin{align}
	S_{\mathrm{NGR}}=m^2_{\mathrm{Pl}}\int |\theta|\left(c_1T_{\mathrm{ax}}+c_2T_{\mathrm{ten}}+c_3T_{\mathrm{vec}}\right),
\end{align}
with $a_1=-\frac{1}{18}c_1+\frac{1}{2}c_2, \ a_2=\frac{1}{9}c_1+\frac{1}{2}c_2, \ a_3=c_3-\frac{1}{2}c_2$, and
\begin{align}
\begin{split}
	&T_\mathrm{vec}=T^\rho{}_{\rho\mu}T_\nu{}^{\nu\mu},
\\
&	T_{\mathrm{ax}}=-\frac{1}{18}\left(T_{\rho\mu\nu}T^{\rho\mu\nu}-2T_{\rho\mu\nu}T^{\mu\rho\nu} \right),\\
	&T_{\mathrm{ten}}=\frac{1}{2}\left(T_{\rho\mu\nu}T^{\rho\mu\nu}+T_{\rho\mu\nu}T^{\mu\rho\nu}\right)-\frac{1}{2}T^\rho{}_{\rho\mu}T_\nu{}^{\nu\mu}\,.
\end{split}
\end{align}
%

%
%
%
%
%
%
%
%
%

\section{Method}

%
%
%
%
%
%
%
%
%

In order to go from the Lagrangian to the Hamiltonian analysis we need to identify the velocities
, derive the conjugate momenta and express everything in canonical variables. We may decompose the torsion scalar in the ADM variables \cite{Okolow:2011np} lapse $\alpha$, shift $\beta^{i}$ and the spatial components of the tetrad $\theta^{A}{}_{i}$:

\begin{align}
	\begin{split}
		\mathbb{T}&=\frac{1}{2\alpha^{2}}T^{A}_{\ i0}T^{B}_{\ j0}M^{i\ j}_{\ A\ B}
		\\&+\frac{1}{\alpha^{2}}T^{A}_{i0}T^{B}_{\ kl}\left[M^{i \ l}_{\ A \ B}\beta^{k}+2\alpha a_{2}h^{il}\xi_{B}\theta^{\ k}_{A}+2\alpha a_{3}h^{il}\xi_{A}\theta^{\ k}_{B} \right]
		\\&+ \frac{1}{\alpha^{2}}T^{A}_{\ ij}T^{B}_{\ kl}\left[\frac{1}{2}M^{\ i\ k}_{\ A\ B}\beta^{j}\beta^{l}+2 \alpha a_{2}h^{jl} \xi_{A}\theta^{\ i}_{B}\beta^{k}+2\alpha a_{3}h^{jl}\xi_{A}\theta^{\ k}_{B}\beta^{i} \right]+{}^{3}\mathbb{T},
	\end{split}
\end{align}
where $h_{ij}=\theta^{A}{}_{i}\theta^{B}{}_{j}\eta_{AB}$ is the induced metric, which is used to raise and lower spatial indices, $
\xi^{A}=-\frac{1}{6}\epsilon^{A}{}_{BCD}\theta^{B}{}_{i}\theta^{C}{}_{j}\theta^{D}{}_{k}\epsilon^{ijk}$ ,
\begin{align}
	M^{i\ j}_{\ A\ B}=-2a_{1}h^{ij}\eta_{AB}+(a_{2}+a_{3})\xi_{A}\xi_{B}h^{ij}-a_{2}\theta^{\ j}_{A}\theta^{\ i}_{B}  -a_{3}\theta^{\ i}_{A}\theta^{\ j}_{B},
\end{align}
and
\begin{align}
	\begin{split}
		{}^{3}\mathbb{T}&\equiv a_{1}\eta_{AB}T^{A}_{\ ij}T^{B}_{\ kl}h^{ik}h^{jl}+a_{2}\eta_{AC}\theta^{C}_{\ m}h^{im}\eta_{BD}\theta^{D}_{\ p}h^{jp}T^{A}_{\ kj}T^{B}_{\ li}h^{kl}
		\\ &+a_{3}\eta_{AC}\theta^{C}_{\ m}h^{im}\eta_{BD}\theta^{D}_{\ p}h^{jp}h^{kl}T^{A}_{\ ki}T^{B}_{lj}.
	\end{split}
\end{align}
Without any loss of generality \cite{Blixt:2018znp}  we can restrict ourselves to the Weitzenböck gauge for which the torsion components are expressed as $T^{A}{}_{\mu\nu}=\partial_{\nu}\theta^{A}{}_{\mu}-\partial_{\mu}\theta^{A}{}_{\nu}$,
and hence the conjugate momenta become,
\begin{align}
	\begin{split}
		\alpha\frac{\pi^{i}_{A}}{\sqrt{h}}&=T^{B}_{\ j0}M^{i\ j}_{\ A\ B}+T^{B}_{\ kl}\left[M^{i \ l}_{\ A \ B}\beta^{k}+2\alpha a_{2}h^{il}\xi_{B}\theta^{\ k}_{A}+2\alpha a_{3}h^{il}\xi_{A}\theta^{\ k}_{B} \right].
	\end{split}
\end{align}
The velocities can now be inverted and expressed in canonical variables using
\begin{align}
	\label{invertvelocity}
	S^{i}_{A}=\dot{\theta}^{B}_{\ j}M^{i \ j}_{\ A \ B},
\end{align}
with
\begin{align}
	\begin{split}
		S^{i}_{A}&=D_{j}\left(\alpha \xi^{B}+\beta^{m}\theta^{B}_{\ m} \right)M^{i\ j}_{\ A\ B}
		\\&-T^{B}_{\ kl}\left[M^{i \ l}_{\ A \ B}\beta^{k}+2\alpha a_{2}h^{il}\xi_{B}\theta^{\ k}_{A}+2\alpha a_{3}h^{il}\xi_{A}\theta^{\ k}_{B} \right]+\alpha \frac{\pi^{i}_{A}}{\sqrt{h}},
	\end{split}
\end{align}
where $D_{i}$ is the Levi-Civita covariant derivative with respect to the induced metric. However, $M$ in equation \eqref{invertvelocity} is singular for certain combinations of parameters of the theory and can hence only be inverted by the Moore-Penrose pseudo-inverse matrix \cite{Ferraro:2016wht}. This is apparent if one decomposes the equation into irreducible representations of the rotation group, which generates the following constraints,
\begin{align}
	\label{D1}
	2a_{1}+a_{2}+a_{3}&=:{}^\mathcal{V}A=0  \implies {}^{\mathcal{V}}C^{i} := S^{i}_{A}\xi^{A}=0,
	\\  2a_{1}-a_{2}&=:{}^\mathcal{A}A=0  \implies {}^{\mathcal{A}}C_{ij}:=S^{k}_{A}\theta^{A}{}_{ [j}h_{i]k}=0,
	\\  2a_{1}+a_{2}&=:{}^\mathcal{S}A=0  \implies {}^{\mathcal{S}}C_{ij}:=S^{k}_{A}\theta^{A}{}_{ (j}h_{i)k}-\frac{1}{3}S^{k}_{A}\theta^{A}_{\ k}h_{ij}=0,
	\\ \label{D4} 2a_{1}+a_{2}+3a_{3}&=:{}^\mathcal{T}A=0  \implies {}^{\mathcal{T}}C:=S^{i}_{A}\theta^{A}_{\ i}=0.
\end{align}
These are primary constraints, since these constrain both the tetrad field and their conjugate momenta, which also can be decomposed into irreducible parts.
In the axial, vector, tensor decomposition we have that
\begin{align}
\label{C1}
c_2+c_3&= {}^\mathcal{V}A=  0,  
\\ \label{C2}  -\frac{2}{9}c_1+\frac{1}{2}c_2&= {}^\mathcal{A}A= 0,  
\\  \label{C3} \frac{3}{2}c_2&= {}^\mathcal{S}A= 0,  
\\ \label{C4} 3c_3&= {}^\mathcal{T}A=0.  
\end{align}
In this language the primary constraints get some further geometrical meaning. Equations \eqref{C1} and \eqref{C2} together imposes the teleparallel equivalent to general relativity and impose invariance of the Lagrangian under pure tetrad local Lorentz transformations\cite{Blagojevic:2000qs}. This is, however, not more apparent from the axial, vector, tensor decomposition we made. What is more interesting are the constraints imposed by equations \eqref{C3} and \eqref{C4}. In this decomposition of the torsion scalar they exactly correspond to putting $T_{\mathrm{ten}}$ and $T_{\mathrm{vec}}$ to zero respectively.
%
%
%
%
%
%
%
%
%

\section{Results}

%
%
%
%
%
%
%
%
%

\begin{figure}[!htb]
	\includegraphics[width=1\textwidth]{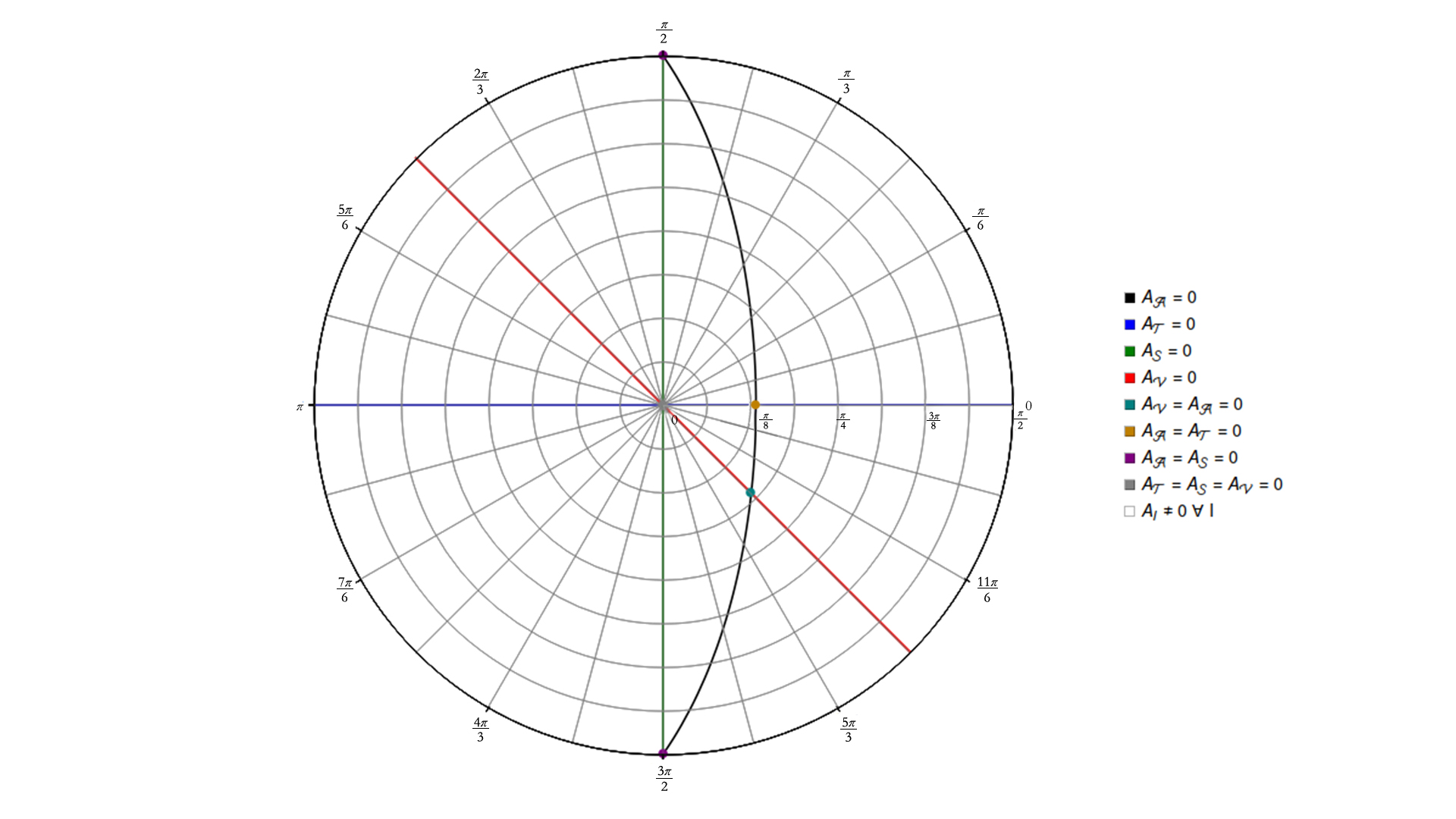}
	\caption{Visualization of the parameter space of new general relativity in coordinates reflecting the axial, vector, tensor decomposition of the Lagrangian, colored by the occurrences of primary constraints. The radial axis shows the zenith angle \(\theta\), while the (circular) polar axis shows the azimuth angle \(\phi\), following the definition~\eqref{eqn:polcoord}.} 
	\label{fig:conplot}
\end{figure}

Different combinations of \eqref{D1}-\eqref{D4} yield 9 non-trivial classes of theories:
\\
\\
\begin{tabular}{|c|c|c|}
\hline
		Theory & Constraints & Location in figure~\ref{fig:conplot} \\ \hline
		$A_{I}\neq 0 \ \forall I\in \{ \mathcal{V},{\mathcal{A}},{\mathcal{S}},\mathcal{T} \}$ &  No constraints & white area
		\\ \hline
		$A_{\mathcal{V}}=0$ &  ${}^{\mathcal{V}}C_{i}=0$ & red line \\ \hline
		$A_{\mathcal{A}}=0$ &   ${}^{\mathcal{A}}C_{ji}=0$ & black line \\ \hline
		$A_{\mathcal{S}}=0$ &   ${}^{\mathcal{S}}C_{ji}=0$ & vertical green line \\ \hline
		$A_{\mathcal{T}}=0$ &   ${}^{\mathcal{T}}C=0$ & horizontal blue line \\ \hline
		$A_{\mathcal{V}}=A_{\mathcal{A}}=0$ & ${}^{\mathcal{V}}C_{i}={}^{\mathcal{A}}C_{ji}=0$ & turquoise point \\ \hline
		$A_{\mathcal{A}}=A_{\mathcal{S}}=0$ &  ${}^{\mathcal{A}}C_{ji}={}^{\mathcal{S}}C_{ji}=0$ & purple points (perimeter) \\ \hline
		$A_{\mathcal{A}}=A_{\mathcal{T}}=0$ &  ${}^{\mathcal{A}}C_{ji}={}^{\mathcal{T}}C=0$ & orange point \\ \hline
		$A_{\mathcal{V}}=A_{\mathcal{S}}=A_{\mathcal{T}}=0$ &  ${}^{\mathcal{V}}C_{i}={}^{\mathcal{S}}C_{ji}={}^{\mathcal{T}}C=0$ & gray point (center) \\ \hline

\end{tabular}
\newline
\newline Any other solutions would be trivial ($c_1=c_2=c_3=0$). Excluding these trivial solutions we can normalize our parameters to
\begin{align}
	\tilde{c}_i=\frac{c_i}{\sqrt{c_1^2+c_2^2+c_3^2}},
\end{align}
for $i=1,2,3$, which means that we can make a 2-dimensional plot to visualize these theories in the normalized parameter-space. This can be nicely visualized in polar coordinates ($\theta,\phi$) on the unit sphere with
\begin{align}
\label{eqn:polcoord}
	\tilde{c}_1=\cos\theta, \quad \tilde{c}_2=\sin\theta\cos\phi, \quad \tilde{c}_3=\sin\theta\sin\phi.
\end{align}
Every pair of antipodal points on the sphere corresponds to a ray in the 3-dimensional parameter space, whose elements describe the same theory. Hence, it suffices to display only the upper half sphere \(\tilde{c}_1 \geq 0\), which is done in figure~\ref{fig:conplot}. However, note that points on the equator \(\tilde{c}_1 = 0\) still appear twice, and both copies should be identified with each other. This applies in particular to the two purple points in figure~\ref{fig:conplot}, both describing the class of theories defined by pure vector torsion \(\tilde{c}_1 = \tilde{c}_2 = 0\).
The Hamiltonian is found to always appear with four Lagrange multipliers (linearity in lapse and shifts) with,
\begin{align}
\begin{split}
\label{generalH}
H&=\alpha\mathcal{H}\left(\theta,M^{-1} \right)+\beta^{k}\mathcal{H}_{k}\left(\theta,M^{-1}\right) + D_{i}\left[\left(\alpha \xi^{A}+\beta^{j}\theta^{A}_{\ j} \right)  \pi^{i}_{A}\right],
\end{split}
\end{align}
in the unconstrained case\cite{Blixt:2018znp}. 

%
%
%
%
%
%
%
%
%

\section{Discussion}

%
%
%
%
%
%
%
%
%

One can distinguish 9 different classes of NGR theories by the presence or absence of primary constraints appearing in their Hamiltonian formulation. What remains to be determined is how many secondary constraints are induced by demanding closure of the constraint algebra. Some considerations in this direction have been studied in \cite{Okolow:2011np,Cheng:1988zg}, however, our work invites for further investigation.
The theories satisfying $A_I\neq0, \forall \ I \in  \{\mathcal{V},\mathcal{A},\mathcal{S},\mathcal{T}\}$ can be parameterized by two free parameters (and a global rescaling of the Lagrangian, fixing the value of the Planck mass, which does not affect the presence or absence of primary constraints). Models which exhibit one primary constraint $A_I=0$ have one free parameter left, while for those with more primary constraints all parameters are fixed. The free parameters might affect the vanishing, or non-vanishing of certain Poisson brackets, which therefore have to be calculated in order to obtain the number of degrees of freedom.
\\ \indent The number of degrees of freedom can be compared with polarization modes in gravitational waves\cite{Hohmann:2018jso}. Furthermore, it can be compared with the linear level in order to find out if the theories are strongly coupled. One may extend this analysis to $f(T_{\mathrm{ax}},T_\mathrm{ten},T_\mathrm{vec})$\cite{Bahamonde:2017wwk} or include parity violating terms.

\end{document}